\RequirePackage{fix-cm}
\documentclass{svjour3}                     
\smartqed  
\usepackage{graphicx}
\usepackage{bm}
\usepackage{xcolor}

\begin{document}

\title{$^{42}$Ca and $^{50}$Ca with the (Many- and Few-body) Unified Method\thanks{This work was
supported by funds provided by the Ministry of Science, Innovation and Universities (Spain) under 
contract No. PGC2018-093636-B-100 .}
}


\author{E. Garrido
        \and A.S. Jensen 
        }


\institute{E. Garrido (corresponding author) \at
              Instituto de Estructura de la Materia, CSIC, Serrano 123, E-28006 Madrid, Spain \\
              \email{e.garrido@csic.es}           
           \and
           A.S. Jensen \at
              Department of Physics and Astronomy, Aarhus University, DK-8000 Aarhus C, Denmark
}

\date{Received: date / Accepted: date}

\maketitle

\begin{abstract}
A new method unifying many and few-body aspects of nuclear
structure has recently been introduced \cite{hov18}.  This method
combines the many-body description of a core and the few-body
structure of this core surrounded by two valence nucleons.  For this
reason this method is expected to work specially well when applied to
nuclei close to the driplines, where the few-body halo structure with
one or more nucleons outside the core is established.  In this work we
apply the new method to nuclei close to the valley of stability, with
$^{42}$Ca and $^{50}$Ca as illustrations.  We compare the results from
uncorrelated mean-field calculations with the ones obtained with the unified 
method allowing arbitrary correlations in the valence space.  We find that the unified
method provides results rather similar, although distinguishable, to
the Hartree-Fock calculations.  The correlations are much less
pronounced than at the driplines, which initially were targets for the
unified method.  The halo structure is not artificially maintained, but
the correlations are here demonstrated to be applicable to well-bound
nuclei.  Excited states built on
valence degrees of freedom are calculated for the same nuclei.
\end{abstract}

\section{Introduction}

An appropriate description of the atomic nucleus, which is a complex
system made of interacting neutrons and protons,
requires a reliable treatment of the corresponding many-body problem \cite{boh69,sie87}.
One of the simplest methods to do so is the use of a mean-field model, where,
by definition,  all correlations are neglected \cite{vau72}. The
opposite approach is the interacting shell model, where all
correlations are included in the theoretical formulation \cite{cau05},
although severely limited in practice.  A large number of different nuclear
models are designed and applied to investigate various aspects of the
nuclear many-body problem \cite{kan01,fel95,hes02,bog03,her16}.  In
particular, few-body methods are intended to study relative motion of frozen
clusters \cite{nie01}, and the no-core shell model to study nuclear
structure materializing at short distance \cite{bru13}.

In classical low-energy nuclear physics the challenge was to account
for both collective and single-particle degrees of freedom in the same
framework \cite{boh69}.  In modern nuclear structure, this challenge
may be similar to incorporating cluster and few-body features in one
model.  This corresponds to different treatment of intrinsic (short
distance) and relative (large distance) cluster degrees of freedom.
Recently, such a unified model was constructed, where the first
formulation uses mean-field approximations and few-body
techniques, respectively, for intrinsic and relative cluster degrees of
freedom \cite{hov18}.  The applications employed are the
Skyrme-Hartree-Fock procedure \cite{vau72} for one cluster (core), and
Faddeev three-body calculations \cite{fad61} for two surrounding (valence)
nucleons.

The traditional few-body techniques have been successfully used to
describe cluster nuclei, where nuclear halos at the driplines are
perhaps the most obvious examples \cite{jen04,zhu93}.  Reliability of
these methods rely on detailed experimental information, which is
necessary to construct the phenomenological potentials between
intrinsically frozen cluster configurations.  The new method has
circumvented these problems, first by allowing the core structure to
adjust to the influence of the valence nucleons, and second, by using the
same nucleon-nucleon interaction between nucleons in the core and
between core and valence nucleons \cite{hov18}.  On the other hand,
the interaction between valence nucleons must be appropriate for the
different Hilbert space allowed for these particles, where variation
from in-medium short-distance to large-distance free interaction is
necessary.

Successful applications of the unified method on dripline nuclei have
already been published. In \cite{hov17} it is applied to $^{26}$O,
which is predicted to show a borromean structure. Also, a good
agreement with the experimental invariant mass spectrum after removal
of one of the valence neutrons is obtained. This invariant mass
spectrum is basically dictated by the derived core-valence neutron
interaction, which, once a nucleon-nucleon interaction has been chosen
to describe the core, is obtained without any additional free
parameter.  In \cite{hov18b} the method is applied in the proton
dripline region. In particular, the two-proton capture reaction
$^{68}\mbox{Se}+p+p\rightarrow ^{70}\mbox{Kr}+\gamma$ is considered,
the corresponding reaction rates computed, and the capture mechanism
investigated.  In \cite{hov18c} the emergence of halos and Efimov
states is investigated by means of the neutron-rich nucleus $^{72}$Ca,
showing how the halo configurations emerge from the mean-field
structure for three-body binding energy less than $\sim100$ keV.

All in all, the unified method has been proved to be efficient in the
description of nuclei close the neutron and proton driplines, i.e., in
the description of nuclear structures containing loosely bound valence
nucleons.  However, an aspect still to be tested is how the method
performs when applied to nuclei in the valley of stability.  At
present the core is treated in the mean-field approximation resulting
in an uncorrelated structure.  Adding two valence nucleons allowed in
the full Hilbert space, except the states occupied by identical core
nucleons, implies that correlations are possible.  The model success
at driplines demonstrates that the weakly bound and spatially extended
halo-like structures are accounted for.

To investigate different possibilities, we choose two stable
well-bound calcium isotopes, $^{42}$Ca and $^{50}$Ca, where each
corresponds to a double magic core ($^{40}$Ca or $^{48}$Ca) plus two
additional well-bound neutrons.  These configurations seem appropriate
as they resemble the cluster structure of a core plus two neutrons
assumed in the unified method.  The simplifications arise from the
empty shells available for the valence nucleons, which simplifies
accounting for the Pauli principle.

The paper is organized with section 2 describing the unified model,
section 3 presents and discusses the ground state results for the two
nuclei, section 4 extends the investigations to excited states of
angular momentum higher than zero, and finally, section 5 contains 
the summary and conclusions.

\section{Theoretical procedure}

The starting point of the method used here is to consider the
many-body system as a clusterized structure. In this way, the natural
choice for the full wave function is the antisymmetrized product of
the wave functions describing each of the clusters and a few-body wave
function describing their relative motion. This method combines then
the many-body description of each individual cluster and the few-body
description associated to the relative motion between clusters.  We
first sketch the method, then focus on the three-body procedure, and
last we consider the crucial issue of the Pauli principle.

\subsection{Sketch of the unified method}

The method is described in detail in Ref.\cite{hov18}, and in here we
shall concentrate on the practical application to a system formed by a
many-body core and two valence nucleons. In this case the full wave
function $\Psi$ can be written as:
\begin{equation}
\Psi={\cal A}\left[  \psi_c(\{\bm{r}_c \})\psi_{3b}(\bm{r}_{cv_1},\bm{r}_{cv_2})
\right],
\label{eq0}
\end{equation}
where ${\cal A}$ is the antisymmetrization operator, $\psi_c$ is the
wave function of the core that depends on a set of spatial and spin
coordinates collected in $\{\bm{r}_c\}$, and $\psi_{3b}$ is the
three-body wave function that depends on the relative coordinates
between the core center of mass and the two valence nucleons,
$\bm{r}_{cv_1}$ and $\bm{r}_{cv_2}$.

In this work the core will be described by means of the 
Skyrme-Hartree-Fock mean field method \cite{vau72}.  Therefore, the core wave
function $\psi_c$ is obtained as the Slater determinant formed by the
single-nucleon wave functions $\phi_i^{q_i}$, where $i$ runs over all
the nucleons in the core and $q_i$ makes explicit the isospin
projection of the nucleon $i$.

The hamiltonian, $H$, and the corresponding interactions are
appropriate for Skyrme-Hartree-Fock calculations.  The equations of
motion are found by minimizing the energy, $E=\langle
\Psi|H|\Psi\rangle$, with respect to simultaneous variation of
single-particle and three-body wave functions, $\phi_i^{q_i}$ and
$\psi_{3b}$, respectively.  The results are a coupled set of
equations, that is
\begin{eqnarray}
\epsilon_i \phi_i^{q_i}(\bm{r})&=&\left[ -\bm{\nabla}\cdot \frac{\hbar^2}{2m_{q_i}^*(\bm{r})} \bm{\nabla}+
U_{q_i}(\bm{r})-i\bm{W}_{q_i}(\bm{r})\cdot \left(\bm{\nabla}\times \bm{\sigma} \right)-\right.  \nonumber \\ & & \left.
-\bm{\nabla}\cdot \frac{\hbar^2}{2m_{q_i}^{\prime*}(\bm{r})} \bm{\nabla}+
U^\prime_{q_i}(\bm{r})-i\bm{W}^\prime_{q_i}(\bm{r})\cdot \left(\bm{\nabla}\times \bm{\sigma} \right)
 \right] \phi_i^{q_i}(\bm{r}),
 \label{eq3}
\end{eqnarray}
\begin{eqnarray}
\lefteqn{
E_{3b} \Psi_{3b}(\bm{x},\bm{y})=}
\label{eq2}  \\ & &
\left[ T_x+T_y+V_{cv_1}(\bm{r}_{cv_1})+V_{cv_2}(\bm{r}_{cv_2})+V_{v_1v_2}(\bm{r}_{v_1v_2})
+V_{cv_1v_2}(\bm{r}_{cv_1},\bm{r}_{cv_2}) \right]\Psi_{3b}(\bm{x},\bm{y}),
\nonumber
\end{eqnarray}
for the core and valence parts, respectively. 

In the Schr\"{o}dinger-like equation Eq.(\ref{eq3}), $\epsilon_i$ is
the single particle energy, $\bm{\sigma}$ is the usual spin operator,
and $U_{q_i}$, $\bm{W}_{q_i}$, and $m_{q_i}^*$ are the
central and spin-orbit mean-filed potentials, and the effective mass function,
respectively.  The primed quantities, $U^\prime_{q_i}$,
$\bm{W}^\prime_{q_i}$, and $m_{q_i}^{\prime*}$ are the part of the
interaction that arises from the valence nucleon densities.  The wave
functions of the valence nucleons necessary to construct
$U^\prime_{q_i}$, $\bm{W}^\prime_{q_i}$, and $m_{q_i}^{\prime*}$ are
obtained from the three-body wave function $\psi_{3b}$, see
Ref. \cite{hov18} for details.

The potentials and effective mass are functions of the parameters of
the chosen Skyrme interaction and the nuclear densities, which are
defined in terms of the single particle functions $\phi_i^{q_i}$
\cite{vau72,cha98}. This is the basis of the well-known
self-consistent Hartree-Fock method, where some trial single-particle
wave functions are used to construct the initial potential and
effective mass functions, such that Eq.(\ref{eq3}) permits to obtain a
new set of single particle functions, which in turn permit to
construct new potentials and effective mass. The procedure is
iteratively repeated til convergence is reached like solving the
usual Hartree-Fock equations \cite{vau72}.

In Eq.(\ref{eq2}), $E_{3b}$ is the three-body energy, $\bm{x}$ and
$\bm{y}$ are the usual Jacobi coordinates, each of them giving rise to
the kinetic energy terms $T_x$ and $T_y$, $V_{cv_i}$ is the
interaction between the core and the valence nucleon $i$, which arises 
from the mean-field potentials obtained from Eq.(\ref{eq3}), and
$V_{v_1v_2}$ is the interaction between the two valence nucleons. The
vectors $\bm{r}_{cv_i}$ and $\bm{r}_{v_1v_2}$ are the relative
coordinate between the core center of mass and the valence nucleon
$i$, and the relative distance between the two valence nucleons,
respectively. The three-body interaction, $V_{cv_1v_2}$, is used for fine-tuning. 
The inclusion of such a term is a common feature in standard three-body calculations, 
where the use of bare two-body interactions typically underbinds the system, \cite{zhu93}.
It is designed to describe effects beyond those of the included pair interactions.
In the present context this means pieces lost by the density dependent parameterization 
of the two-body Skyrme interaction.  In few-body physics, it accounts for three-body effects 
of cluster deformation or polarization.  Provided that this three-body force has short-range 
character, its role is basically to  shift the energy while maintaining the structure of the system.

The self-consistent practical procedure is now simple. In the initial
step we solve the Hartree-Fock equations (\ref{eq3}) for the core and
obtain an initial core-valence nucleon interaction, which is used in
Eq.(\ref{eq2}) to obtain an initial three-body solution
$\psi_{3b}$. This wave function is used to construct the valence
nucleon single-particle wave functions used in Eq.(\ref{eq3}) in order
to get a new core-valence nucleon interaction, which in turn is
used again in Eq.(\ref{eq2}) to get a new three-body wave function
and new valence nucleon single-particle wave functions to be
used in Eq.(\ref{eq3}).  The procedure using the output of
Eq.(\ref{eq2}) as input for Eq.(\ref{eq3}), and the output of
Eq.(\ref{eq3}) as input for Eq.(\ref{eq2}) is iterated til convergence
is reached.

Both, core and valence structure are influencing each other.  The core
is only marginally disturbed by the two valence nucleons but still
often amounting to a few hundred keV.  On the other hand, the valence
nucleons are completely controlled by the interaction from the nucleons in
the core.

\subsection{The three-body calculation}

In this work the three-body problem given by the Schr\"{o}dinger
equation (\ref{eq2}) will be solved, not actually by solving directly
the equation itself, but by solving the Faddeev equations, whose sum
provides the Schr\"{o}dinger equation (\ref{eq2}), but which permits a
democratic treatment of the internal two-body subsystems. The
three-body wave function is written as a sum of three components, each
of them depending on each of the three possible sets of Jacobi
coordinates, in such a way that each of the three Faddeev equations
contains one of the three two-body potentials in its natural
coordinate \cite{fad61}.

To be more precise, we shall solve the Faddeev equations in coordinate
space by means of the hyperspherical adiabatic expansion method
described in \cite{nie01}. In this method the total three-body wave
function is written in terms of the usual hyperspherical coordinates
(one radial coordinate, i.e., the hyperradius, and five hyperangles),
and it is computed in a two-step procedure. In the first step the
hyperradius is considered as a parameter, and the angular eigenvalue
problem is solved for fixed values of the hyperradius. The angular
eigenfunctions are used as a complete basis set in order to expand the
three-body wave function. The radial coefficients in this expansion (functions of the
hyperradius) are obtained in a second step as a solution of a coupled
set of differential radial equations, where the angular eigenvalues 
 enter as effective (adiabatic)
potentials. See Ref.\cite{nie01} for details.
 
At this point it is important to mention that the effective mass term,
$\hbar^2/(2m_{q_i}^*)$, contained in Eq.(\ref{eq3})
takes the form $\hbar^2/(2m_{q_i})-n_a(\bm{r})$, where $m_{q_i}$ is
the constant nucleon mass, and the explicit form of the function $n_a$
can be found for instance in \cite{vau72}. This fact implies that the
interaction $V_{cv}$ between the core and the valence nucleons
contains not only the central and spin-orbit parts $U_{q_i}$ and
$\bm{W}_{q_i}$, but also a gradient term of the form $\bm{\nabla}\cdot
n_a(\bm{r}) \bm{\nabla}$. The presence of this term complicates a
little bit the calculation of the three-body wave function.  In the
particular case of using the adiabatic expansion method, the gradient
term in the $V_{cv}$ potential gives rise to a family of new coupling
functions in the set of differential radial equations, not described
in Ref.\cite{nie01}, but derived in detail in Appendix B of
\cite{hov18}.

\subsection{The Pauli principle}

The unified method sketched above contains the underlying assumption
that the nucleons in the core occupy the lowest possible states, in
such a way that only the valence nucleons can move between different
orbits. The mean-field calculation used to describe the core
guaranties the orthogonality between the different occupied orbits,
preventing identical fermions from occupying the same single-particle
state in the core.

However, the core-valence nucleon interaction $V_{cv_i}$ generated by
the core mean-field is such that the two-body system formed by the
core and the valence nucleon shows a series of bound states, each of
them with its corresponding $\ell_j$ quantum numbers, i.e., orbital
angular momentum $\ell$ and total angular momentum $j$. The important
point is that this series of bound two-body states contains all the
$\ell_j$-states corresponding to the nucleons in the core, therefore
already occupied, and forbidden to the valence nucleons by the Pauli
principle. It is then clear that these states must be excluded from
the space employed when solving the three-body equation (\ref{eq2}).

An efficient method to exclude the Pauli forbidden states in the
three-body calculation is the use of, not $V_{cv_i}$ as the
core-valence nucleon interaction, but of its phase-equivalent version
\cite{gar99}. This amounts to construct a new potential having exactly
the same phase shifts as $V_{cv_i}$ for all energies, but whose bound
state two-body spectrum does not contain the state(s) forbidden by the
Pauli principle.

In the case of using the adiabatic expansion method, a second procedure
can also be used in order to exclude the Pauli forbidden states.  This
method exploits the fact that the effective adiabatic potentials
entering in the radial equations are asymptotically associated to
specific two-body structures. In particular, some of them will be
associated to the Pauli forbidden bound two-body states.  The
procedure is then simply to exclude these terms from the adiabatic
expansion, or, in other words, to exclude the radial differential
equations containing the unwanted adiabatic potentials from the set to
be solved when computing the radial wave functions \cite{gar97}. This
method is particularly efficient when the adiabatic components to be
excluded are, to a large extent, decoupled from the rest. This happens
very often with the very deep Pauli forbidden states.

 In this work the Pauli principle will be implemented by use of a combination 
 of the two methods described above. If the adiabatic channel
associated to a given Pauli forbidden state is very much decoupled
from the rest, the second method will be used. Otherwise, we shall
construct the corresponding phase equivalent potential.
The antisymmetry between core nucleons is then implemented through the
Slater determinant, and between valence nucleons directly by using
only two-body configurations of the allowed symmetry.  
But also, when computing the distorted mean field due to the presence 
of the valence nucleons, the wave functions of the valence
nucleons are antisymmetrized as well with respect to the ones of the nucleons
in the core. In other words, across core and valence space, only antisymmetric 
matrix elements are used in the calculations.
Furthermore, only core-unoccupied orbits are allowed for identical
valence nucleons.

\section{The 0$^+$ ground state of $^{42}$Ca and $^{50}$Ca}

We use the unified method to compute the 0$^+$ ground state of
$^{42}$Ca and $^{50}$Ca as three-body systems of two neutrons
surrounding cores of $^{40}$Ca and $^{48}$Ca, respectively.  The
self-consistent mean-field calculations provide a sequence of
core-occupied single-particle neutron states, that is, with the 
shell-model notation, the $1s_{1/2}$, $1p_{3/2}$, $1p_{1/2}$, $1d_{5/2}$,
$2s_{1/2}$, and $1d_{3/2}$ states, which amounts to the 20 states occupied 
by the neutrons in $^{40}$Ca.  These states are Pauli forbidden to the
two valence neutrons in $^{42}$Ca, which are therefore pushed to the 
next states, i.e.,  the $1f_{7/2}$, $2p_{3/2}$, or higher states.  
In $^{48}$Ca also the eight $1f_{7/2}$ neutron states are occupied, and the valence 
neutrons in $^{50}$Ca are pushed to the $2p_{3/2}$ or higher states.

\subsection{Effective three-body potentials}

\begin{figure}
\begin{center}
\includegraphics[width=8cm,angle=0]{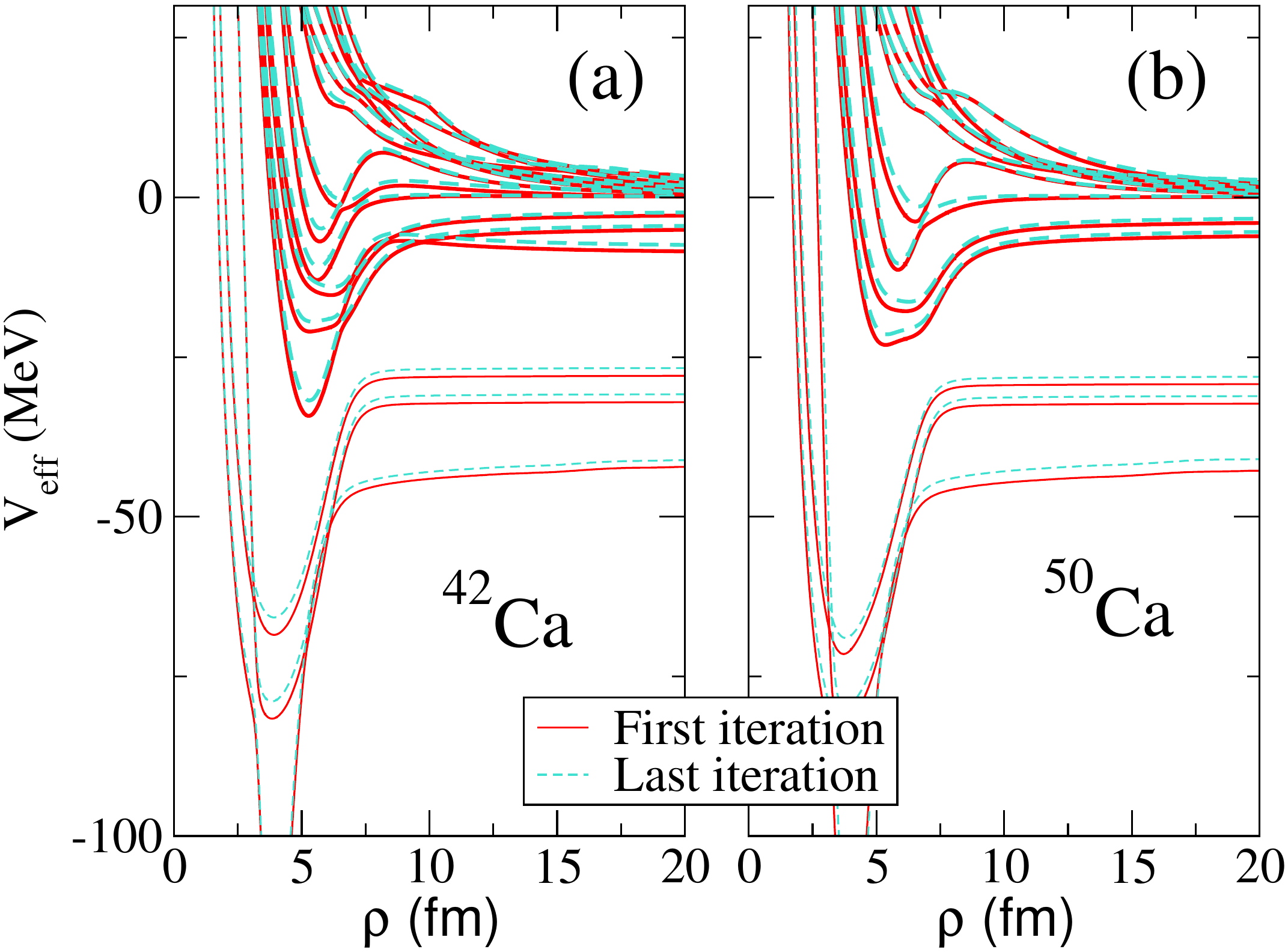}
\end{center}
\caption{For $^{42}$Ca, panel (a), and $^{50}$Ca, panel (b), three-body effective adiabatic potentials obtained after description of the 
core by means of the SKM$^*$ Skyrme interaction. The solid and dashed curves are the potentials obtained after the first 
iteration and once the convergence has been reached, respectively. The thin curves correspond to the adiabatic potentials
associated to the $1s_{1/2}$, $1p_{3/2}$, and $1p_{1/2}$ Pauli forbidden states that have not been eliminated by means 
of the phase equivalent potentials.}
\label{fig1} 
\end{figure}

The unified method leaves no freedom in the choice of the core-valence
nucleon interaction, and as a consequence, the adiabatic potentials
entering in the three-body calculation are completely dictated by the
chosen effective nucleon-nucleon interaction used to describe the
core.  The three-body calculations employ the hyperspherical adiabatic
expansion method, where we include all partial waves with relative
orbital angular momenta up to $\ell=4$.  This amounts to 5 partial 
waves in the (first) Jacobi set with $\bf{x}$ between the two valence 
neutrons, and 9 partial waves in other two (second and third) identical 
Jacobi sets. In Fig.~\ref{fig1} we show
the effective adiabatic potentials for the 0$^+$ ground state in
$^{42}$Ca and $^{50}$Ca when the SKM$^*$ Skyrme interaction is chosen.
The structures are already established after the first iteration (solid
curves), which are very similar, although distinguishable, to the converged
potentials (dashed curves).

The Pauli forbidden core states have been excluded by use of the
corresponding phase equivalent potentials, except for the deep bound
states in the $s$ and $p$-shells.  They are seen in Fig.~\ref{fig1} as
the three deep states (thin curves) approaching asymptotically at
large distance the two-body neutron-core energies around $-43$~MeV,
$-32$~MeV, and $-28$~MeV corresponding to the $1s_{1/2}$, $1p_{3/2}$,
and $1p_{1/2}$ states.  These potentials are, for both nuclei, clearly
decoupled from the higher-lying potentials, and they are directly omitted
from the subsequent three-body calculation.

Among the remaining potentials (thick curves) in Fig.~\ref{fig1}a,
still three of them asymptotically approach negative energies at large
distance.  They correspond to the empty $1f_{7/2}$, $2p_{3/2}$, and
$2p_{1/2}$ shells, which therefore are the most likely to be used by
the two valence neutrons.  For $^{50}$Ca in Fig.~\ref{fig1}b, only two
of them are left with asymptotic negative energies.  The third one,
corresponding to the $1f_{7/2}$ shell, is now fully occupied by the
neutrons in the core, and it has therefore been removed by means of the
corresponding phase equivalent potential.

\begin{figure}
\begin{center}
\includegraphics[width=8cm,angle=0]{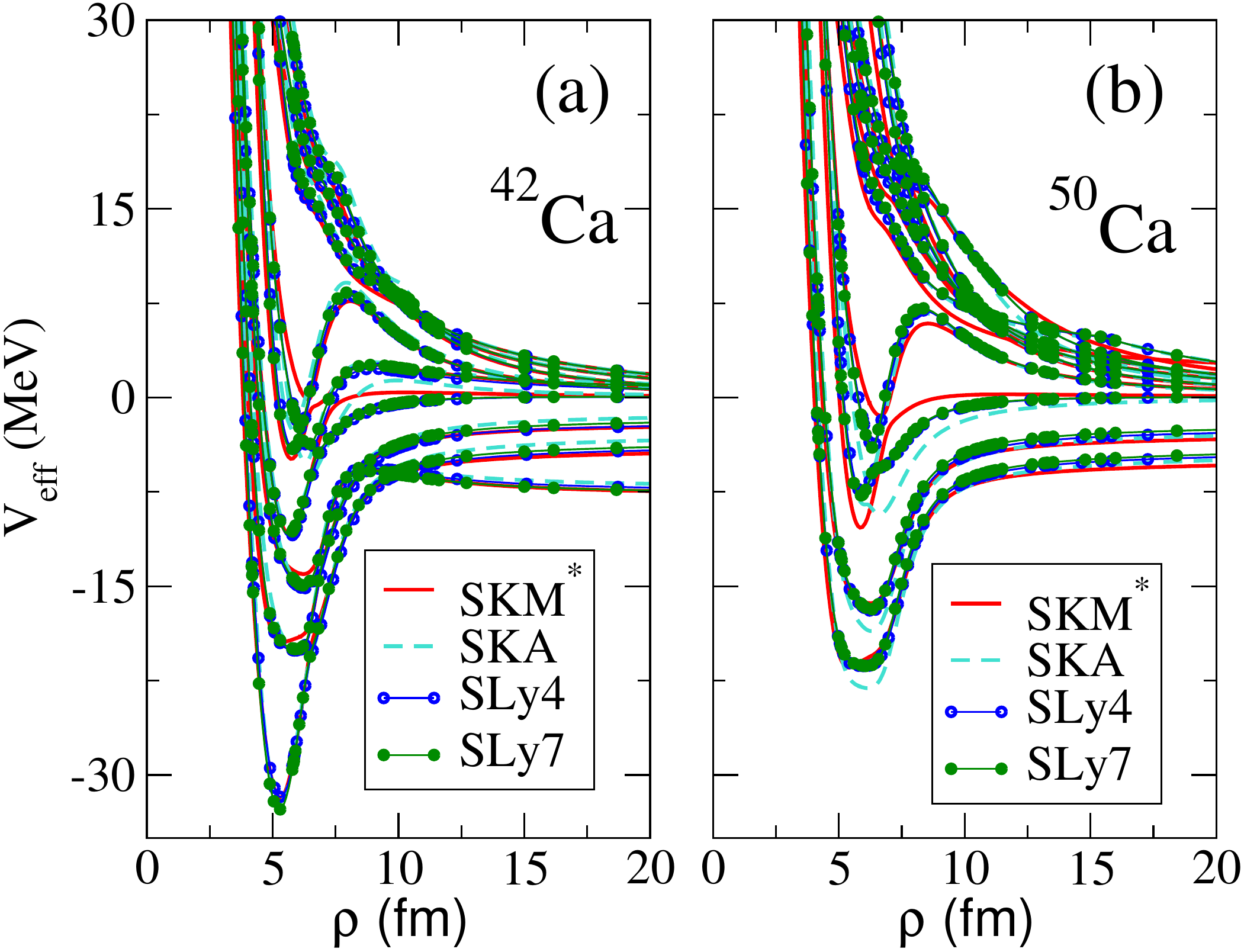}
\end{center}
\caption{For $^{42}$Ca, panel (a), and $^{50}$Ca, panel (b), converged
  three-body effective adiabatic potentials obtained after description
  of the core by means of the SKM$^*$, SKA, SLy4, and SLy7 Skyrme
  interactions \cite{cha98}.}
\label{fig2} 
\end{figure}

A number of effective interactions are available even for use within
the mean-field approximation.  To study such uncertainties, we compare
in Fig.~\ref{fig2} the decisive adiabatic potentials obtained with
four different Skyrme interactions \cite{cha98}.  The differences are
minor, and therefore the dependence on the Skyrme interaction is not
expected to be significant.  We have also compared results from two
different neutron-neutron interactions for use entirely
within the valence space, the simple gaussian potential
given in \cite{gar97}, and the more sophisticated Argonne $v_{18}$
potential described in \cite{wir95}.  The results again are only marginally
different, that is, less $150$~keV in the final two-neutron separation
energies.

\begin{figure}
\begin{center}
\includegraphics[width=11cm,angle=0]{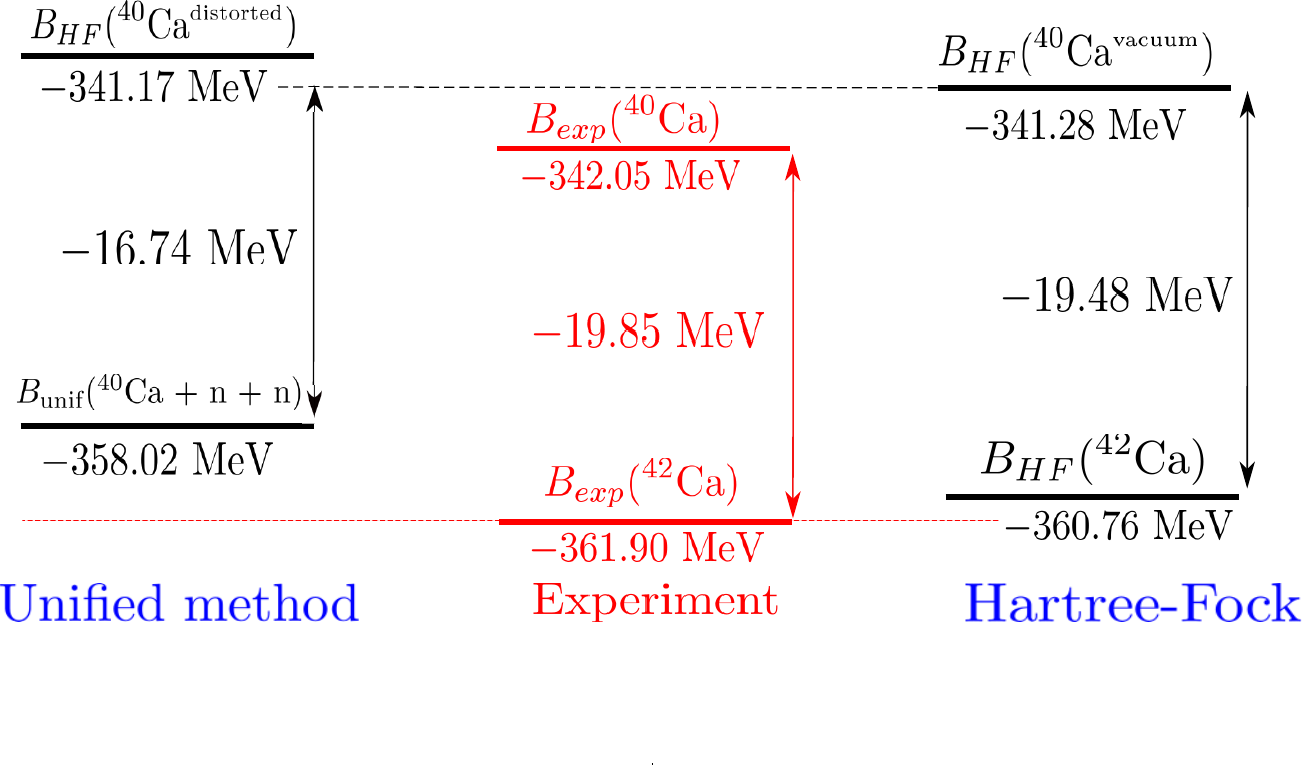}
\end{center}
\caption{$^{40}$Ca and $^{42}$Ca binding energies obtained with the unified method (left) and the Hartree-Fock method (right). The SKM$^*$ 
interaction has been used. The experimental values are given in the central part \cite{fir99}. }
\label{fig3} 
\end{figure}

\subsection{Energies}

The results with the unified method can immediately be compared with
pure Hartree-Fock calculations, which are expected to work well for
$^{42}$Ca and $^{50}$Ca, simply because the effective interactions are
adjusted to achieve that goal on average for a number of spherical
nuclei.  The right part of Fig.~\ref{fig3} gives the binding energies of
$^{42}$Ca and $^{40}$Ca obtained in a Hartree-Fock calculation with
the SKM$^*$ interaction.  The difference between these energies is the
Hartree-Fock two-neutron separation energy of $-19.48$ MeV.  These
results agree reasonably well with the experimental values in the
central part of the figure \cite{fir99}.  Both energies are apparently shifted
towards less binding by roughly the same amount leaving the
two-neutron separation energy almost at the measured value.

The left part of the figure gives the results obtained with the method
discussed in this work.  This means that the $^{40}$Ca core is
described after a Hartree-Fock calculation under the presence of the
two valence neutrons.  This presence is distorting the mean field,
which makes the $^{40}$Ca core actually slightly less bound than the
isolated $^{40}$Ca nucleus, due to the distortion away from the
optimal configuration.  After convergence, we obtain a $^{42}$Ca
nucleus 2.74~MeV less bound than the pure Hartree-Fock result, leading
to the same difference in the two-neutron separation energy.  Note
that this two-neutron separation energy is taken as the binding energy
difference between the computed $^{42}$Ca nucleus and the ``isolated''
$^{40}$Ca nucleus.  Use of a different Skyrme interaction does not
significantly modify the results shown in Fig.~\ref{fig3}.
The same energies for $^{48}$Ca and $^{50}$Ca reveal the same
qualitative behavior, namely with Hartree-Fock results in reasonable
agreement with the experimental values, and with two-neutron separation
energies in the vicinity of $-11$ MeV.  Again the unified method gives
underbinding for $^{50}$Ca by between $0.7$~MeV and $2$~MeV,
depending on the Skyrme force used,  compared to the
Hartree-Fock calculations.

\begin{table}
\caption{Two-neutron separation energies, $S_{2n}$, in MeV for $^{42}$Ca and $^{50}$Ca, and four different Skyrme interactions.  
The labels ``Zero'' and ``Conv.'' refer to our calculation after the initial iteration and after reaching convergence, respectively.
The ``HF'' label refers to the Hartree-Fock calculation. The results within parenthesis are obtained after inclusion 
of BCS.  The experimental data are taken from \cite{fir99}. }
\label{tab1}       
\begin{center}
\begin{tabular}{ll|cccc}
\hline\noalign{\smallskip}
  &    &  \multicolumn{4}{c}{$S_{2n}$ (MeV)}    \\
  &    &  Zero  &Conv.   &   HF (+BCS)   &  Exp. \cite{fir99}  \\
  \noalign{\smallskip}\hline\noalign{\smallskip}
  $^{42}$Ca  &    SKM$^*$  &$-18.70$ &$-16.74$   &  $-19.48 (-21.94)$ & $-19.84$     \\
                     &    SLy4          & $-17.89$ &$-16.30$  &  $-18.29 (-20.25)$   &       ``        \\
                     &    SKa          &   $-16.99$ &$-15.68$ &   $-17.70 (-20.01)$   &       ``       \\      
                    &    SLy7          &   $-18.29$ &$-16.66$&   $-18.78 (-21.11)$ &         ``          \\
                    \noalign{\smallskip}\hline\noalign{\smallskip}
  $^{50}$Ca  &    SKM$^*$  & $-11.99$&$-10.85$&  $-12.48 (-13.16)$  &  $-11.50$  \\
                     &    SLy4          &  $-10.58$&$-9.88$  &  $-10.58 (-11.27)$  &    ``             \\
                     &    SKa           & $-11.00$ & $-10.41$&  $-10.90 (-11.58)$  &      ``          \\  
                    &    SLy7          &  $-10.12$ &$-9.51$  &   $-10.07 (-10.73)$ &      ``            \\
\noalign{\smallskip}\hline
\end{tabular}
\end{center}
\end{table}

The summary of the results for the two-neutron separation energies are given in
Table~\ref{tab1} for several Skyrme interactions. Under the labels ``Zero'' and
``Conv.'' we give, respectively, the computed two-neutron separation energies after the initial
iteration and after convergence has been reached.  As we can see, 
right after the first iteration the computed energies are not 
very different to the ones obtained after a Hartree-Fock calculation (given
in the table under the label ``HF''). However, as shown in Fig.~\ref{fig1},
the iteration procedure, which accounts for the distortion of the core due
to the presence of the valence neutrons, gives rise to less deep adiabatic potentials,
in such a way that the system becomes less bound than after a Hartree-Fock 
calculation, as illustrated in Fig.~\ref{fig3}.
 
In the table we also give the results obtained after a Hartree-Fock
plus BCS calculation, which improves the agreement with the
experimental values.  This is not surprising  since the pairing strength is 
chosen phenomenologically to reproduce odd-even differences in average 
binding energies. However, the unified method does not pick up this 
pairing effect in energy.

This fact suggests that our interaction between the 
valence neutrons is too weak in the pairing channel to reproduce the experimental results. 
The reason can very likely be that the interaction between the valence neutrons used in our calculation is just 
a free space neutron-neutron interaction \cite{gar97} (which, as mentioned, provides results very similar
to the ones obtained with the Argonne potential  \cite{wir95}). This is necessary with the 
non-restricted valence Hilbert space, since otherwise the Skyrme zero-range interactions
would lead to undesired divergences \cite{hov18}. The price to pay for this is 
a poor description of the off-shell (mainly three-body) character of the valence neutrons. Obviously,
for nuclei close to the driplines, where the valence neutrons are not that close to the core,
the inaccuracy introduced by the use of a bare neutron-neutron interaction is much
less pronounced. 

Another point to mention here is the effect of the use of phase equivalent
potentials in order to eliminate the highest-lying mean-field neutron states
forbidden by the Pauli principle. As mentioned, the scattering properties
of the new potential are identical to the ones of the original potential,
but it has the effect of, first,  introducing a repulsion at short distances, and,
second, removing one node at small distance in the wave functions  available to 
the valence neutron. This effect, which is also reducing the computed
two-neutron separation energy, is clearly largest when the densities at
small distances are significant, and, again, it is expected to play a minor
role when dealing with nuclei close to the driplines.

Summarizing, in the description of well-bound nuclei close to the valley
of stability, since the two valence neutrons are relatively close to the core,
the use of a free-space neutron-neutron interaction and the use of phase 
equivalent potentials to take care of the Pauli principle, give rise to a 
non-negligible underbinding of the system.  Close to the driplines, where
the valence neutrons are further apart, this problem is much less important. 
In any case, the underbinding shown in Table~\ref{tab1}, although $10 - 15\%$ in
two-neutron separation energy, is less than 1\% in the total binding energy.

\subsection{Relative sizes}

An indication that the energy differences in Table~\ref{tab1}
between our method and the Hartree-Fock method are rather modest,
is the fact that the effective three-body force required to fit the 
experimental value is weak enough to keep the structure of the system 
essentially unchanged.

\begin{table}
\caption{For four different Skyrme interactions, we give, in fm, the mass root-mean square radii, $r_{rms}$, of the
core, $^{40}$Ca and $^{48}$Ca, of the full nucleus, $^{42}$Ca and $^{50}$Ca, and the root-mean square distances between the 
core and a valence neutron, $r_{cn}$, and between the two valence neutrons
$r_{nn}$. The label ``Comp.''  refers to our calculation using the unified method, and ``Comp.+3bd'' refers to our calculation including an effective 
three-body force such that the experimental two-neutron separation energy is reproduced. The label ``HF''' indicates a Hartree-Fock calculation.}
\label{tab2}       
\begin{center}
\begin{tabular}{ll|cccc|cccc}
\hline\noalign{\smallskip}
&  & \multicolumn{4}{c|}{$^{42}$Ca} &  \multicolumn{4}{c}{$^{50}$Ca} \\
&  &  SKM$^*$  &  SLy4  &  SKa  &   SLy7 &  SKM$^*$  &  SLy4  &  SKa  &   SLy7 \\ 
\noalign{\smallskip}\hline
\noalign{\smallskip}
$r_{rms},$& Core &  3.40  &  3.40  &  3.39  &  3.37  &  3.55  &  3.56  &   3.56 & 3.54\\
\noalign{\smallskip}\hline
\noalign{\smallskip}
$r_{rms}$,& Comp.          &  3.44  &  3.44  &  3.44  &  3.44  &  3.57  &  3.58  &   3.58 & 3.58\\
                 & Comp.+3bd &  3.43  &  3.43  &  3.43  &  3.43  &  3.57  &  3.57  &   3.57 & 3.57\\
                 & HF               &  3.44  &  3.44  &  3.43  &  3.41  &  3.60  &  3.61  &   3.62 & 3.59\\
\noalign{\smallskip}\hline
\noalign{\smallskip}
$r_{cn}$,& Comp.          &  4.12  &  4.15  &  4.18  &  4.12  &  4.50  &  4.59  &   4.61 & 4.58\\
               & Comp.+3bd &  4.09  &  4.09  &  4.10  &  4.07  &  4.46  &  4.53  &   4.54 & 4.46\\
               & HF               &  4.06  &  4.10  &  4.13  &  4.07  &  4.47  &  4.60  &   4.66 & 4.60\\
\noalign{\smallskip}\hline
\noalign{\smallskip}
$r_{nn}$,& Comp.          &  5.54  &  5.58  &  5.61  &  5.55  &  6.02  &  6.06  &   6.05 & 6.04\\
               & Comp.+3bd &   5.51 &  5.50  &   5.52 &  5.49  &   5.97 &  6.03  &   5.98 & 5.95\\
\noalign{\smallskip}\hline

\end{tabular}
\end{center}
\end{table}

The simplest observable revealing structure are the root-mean square radii, which
give a measure of the size and the geometry of the system. For the two 
nuclei and the four Skyrme interactions used in this work, we
give in Table~\ref{tab2} several of these second moment expectation
values of size measures, that is, the total mass root-mean-square radii ($r_{rms}$)
for the core and the $^{42}$Ca and $^{50}$Ca nuclei, the distance between the core center-of-mass
and one of the valence neutrons ($r_{cn}$), and  the distance between
the two valence neutrons ($r_{nn}$).  The label ``Comp.+3bd'' refers to the result obtained with
our method when a three-body force is used to fit the experimental
two-neutron separation energy.

The first overall conclusion is that the calculations including the
three-body potential are almost identical to those without this
force. This is no surprise, rather the intention, by choosing a
structureless dependence of the three-body force on only the 
hyperspherical radius. The
combination with rather strong three-body binding results in an
approximately constant shift of the potential around its minimum.

The total rms values are not very sensitive to the addition of two valence
neutrons, since they are essentially determined by the Hartree-Fock core
structure.  Comparing $r_{cn}$ in Table~\ref{tab2} from the two
methods, we see that the effect of the three-body potential is a small
decrease of these radii, which bring them very close to the
Hartree-Fock result.  These results show that this marginal effect most
probably arises from the lowering of the energy, that is moving away
from the threshold of no binding.

The average distance between valence neutrons is given by $r_{nn}$.
These radii can be compared to the average distance between nucleons
in the core, $r_{cc} \approx r_{rms}\sqrt{2}$, which is roughly the same in the Hartree-Fock
treatments of the core and the total system for both $^{42}$Ca and $^{50}$Ca.
The actual core sizes of $3.44 (3.57) \sqrt{2} = 4.85 (5.03)$ fm are
then about $10\% (15\%)$ smaller than the computed valence neutron-neutron distances.

The average distance in the unified model from one valence neutron to
one nucleon in the core can be estimated as the expectation value of
$(\bm{r}_{nc} - \bm{r})^2$ over the core distribution with
coordinate $\bm{r}$. For a very heavy core we then get $\sqrt{r_{cn}^2
  + r_{rms}^2}$ (since the cross term vanishes after integration), which
is, respectively, about $5.3$ fm and $5.7$ fm. Not surprisingly, these 
values are between the values of $r_{cc}$ and $r_{nn}$ for the two
nuclei.

These results are consistent with previous results for dripline nuclei
\cite{hov17,hov18c}, which is also a reflection of the less
restricted Hilbert space in the unified method than in mean-field
calculations.  This in turn allows the valence neutrons to benefit
fully from the largest attraction arising from the core by being far
apart if required, perhaps even on opposite sides of the core.  The
structures in the two methods are very similar but with relatively
small differences in striking contrast to the relative sizes found
very close to driplines with the halo characteristics.  We emphasize
that the cluster structure assumed a priory in the unified model has
sufficient flexibility to describe continuously uncorrelated
mean-field as well as halo structures \cite{hov18,hov18c}.

\begin{table}
\caption{ Weight (in \%) of the most contributing components to the norm of the $^{42}$Ca and $^{50}$Ca three-body wave functions. In the notation
(${\ell_x}_{j_x}$,  ${\ell_y}_{j_y}$) the quantum numbers $\ell_x$ and $\ell_y$ are the relative orbital angular momenta between the core and one
of the valence neutrons, and between the center of mass of the core-neutron system and second valence neutron, respectively.  The coupling between $\ell_x$ ($\ell_y$) 
and the spin of the first (second) valence neutron provides the angular momentum $j_x$ ($j_y$). }
\label{tab3}       
\begin{center}
\begin{tabular}{l|cccc}
\hline\noalign{\smallskip}
  &    \multicolumn{4}{c}{$^{42}$Ca}   \\ 
  &    $(f_{7/2}, f_{7/2})$ &  $(f_{5/2}, f_{5/2})$ &$(p_{3/2}, p_{3/2})$ &  $(p_{1/2}, p_{1/2})$ \\\noalign{\smallskip}
                                                                                                                                                                        \hline                                                                                                                                                                        \noalign{\smallskip}
  SKM$^*$ &    93.7 &   2.8 &  1.1  & 1.1 \\
  SLy4         &    93.8 &   2.6 &  1.1 &  1.1 \\
  SKa           &    94.2 &   2.5 &  0.9 &  0.9 \\
  SLy7         &    94.4 &   2.5 & 1.0 & 0.9 \\\noalign{\smallskip}
  \hline\noalign{\smallskip}
  &     \multicolumn{4}{c}{$^{50}$Ca}     \\ 
  &    $(p_{3/2}, p_{3/2})$   &    $(p_{1/2}, p_{1/2})$    & $(d_{5/2}, d_{5/2})$  &  $(g_{9/2}, g_{9/2})$ \\\noalign{\smallskip}
                                                                                                                                                                        \hline                                                                                                                                                                        \noalign{\smallskip}
  SKM$^*$ &    95.0    &   3.2  & 0.3  & 0.5\\
  SLy4         &    94.7    &   3.5  & 0.4 & 0.2 \\
  SKa           &    94.6    &   3.6  & 0.3 & 0.2 \\
  SLy7         &    94.7   &   3.5  & 0.4  &  0.2\\

\noalign{\smallskip}\hline
\end{tabular}
\end{center}
\end{table}

To further analyze the structure we calculate the partial wave
decomposition, shown in Table~\ref{tab3}, for the three-body ground
state wave functions of  $^{42}$Ca and $^{50}$Ca.  Let us take $\ell_x$
as the relative orbital angular momentum between the core and one of
the valence neutrons, and $j_x$ as the coupling between $\ell_x$ and
the spin of the neutron.  In the same way, we take $\ell_y$ as the
relative angular momentum between the center of mass of the
core-neutron system and second valence neutron, which after coupling
to the spin of the second neutron gives the angular momentum $j_y$.

The two valence neutrons are in pairs of time reversed single-particle
states.  The weights of the four most contributing (${\ell_x}_{j_x}$,
${\ell_y}_{j_y}$) components to the total norm are given
for the ground state wave functions of $^{42}$Ca
and $^{50}$Ca.  As expected, for $^{42}$Ca the two valence neutrons
mainly occupy the $f_{7/2}$ shell, although the three-body calculation
predicts small contributions from other two-neutron components, $\sim
2.8 \%$ provided by $f_{5/2}$ shell and $\sim 2 \%$ by the $p$
shells.  The same happens for $^{50}$Ca, where the expected structure
with the two valence neutrons are moved to higher (but lowest
available) shells, that is clearly the $p_{3/2}$ shell dominates with
$\sim 95$\%, and contributions of $\sim 3 \%$, $\sim 0.35 \%$, $\sim
0.3 \%$ provided by the shells $p_{1/2}$, $d_{5/2}$ and $g_{9/2}$,
respectively.

The linear combination of different configurations is an extension of
the Hilbert space used in Hartree-Fock calculations. These admixtures
describe correlations beyond mean-field approximations in the few-body
wave function and consequently in the total unified wave function.
For $^{42}$Ca the $f_{5/2}$ states are more favorable than $p$-states,
that is the reversed order compared to the shell model and ordinary
Hartree-Fock mean field calculations.  This is a signature of the
pairing correlation where higher angular momenta are preferred.  The
arbitrary valence correlations allowed in the unified method is here
picking up the pair structure described in the BCS-approximation.  The
$d$ and $g$-states from the next major shell are not present,
apparently because they would cost too much in energy.

For $^{50}$Ca the $f_{5/2}$ state is not contributing although next in
the shell model sequence and therefore it should be directly energetic
favorable.  In contrast, the higher lying (in shell model) $d$ and
$g$-states give visible contributions of the BCS-type of correlations.
All these small contributions are significant, at least qualitatively,
although most likely sensitive to the angular dependence of the
interaction between the valence neutrons.

\section{Excited states of $^{42}$Ca and $^{50}$Ca}

The method used here permits as well to compute the excited three-body
states arising from the excitation of the valence neutrons.  The
present method assumes a $0^+$ structure for the core, and the total
angular momentum has then to be carried entirely by the valence
particles.  This means that applications are limited to sufficiently
low total energies, at least lower than the lowest core excitation energy
(3.4 MeV and 3.8 MeV for $^{40}$Ca and $^{48}$Ca, respectively \cite{fir99}), where couplings 
to core excited states are absent or negligibly small.

The $j$ shells mainly occupied by the two valence neutrons can support
even angular momentum states from ground state $0^+$ to $(2j-1)^+$.
We investigate the applicability of the unified method to such states
by calculations where the three-body space is restricted to the
desired total angular momentum \cite{nie01}.  The dependence on the
choice of the Skyrme interaction is very weak for the ground states and we
therefore only use one of these interactions, SKM$^*$.  We choose again
the nuclei $^{42}$Ca and $^{50}$Ca, and compute excited states with non-zero
angular momenta $J^\pi$.

An important difference compared to the $0^+$ case is that the number of partial waves 
coupling to the selected total angular momentum is now clearly higher. For instance, for
the $2^+$ states, if relative angular momenta between the particles up to $\ell=4$ 
are included, we get 15 components in the first Jacobi set (5 in the 0$^+$ case), and 31 in the 
other two sets (9 in the 0$^+$ case). 
This results in much heavier numerical calculations. However, we have observed that
the lowest $2^+$ states are highly dominated, even by more than 98\% of the norm, by the 
$f$- and $p$-wave relative momenta
between the core and the valence neutrons for $^{42}$Ca and $^{50}$Ca,
respectively. For this reason, in order to simplify the calculations, the results shown in this section have been
obtained including only core-neutron relative $f$-waves for $^{42}$Ca, 
and only relative $p$-waves for $^{50}$Ca.

\begin{table}
\caption{Computed, $E_{comp}^*$, and experimental, $E_{exp}$, excitation energies, in MeV,  of the $2^+$, $4^+$, and $6^+$ excited states in $^{42}$Ca, and the $2^+$ excited
state in $^{50}$Ca, arising from the excitation of the valence neutrons. The experimental values are taken
from \cite{fir99}.}
\label{tab4}       
\begin{center}
\begin{tabular}{l|ccc|c}
\hline\noalign{\smallskip}
  &   \multicolumn{3}{c|}{$^{42}$Ca} &  $^{50}$Ca    \\
  &  2$^+$  &   4$^+$  &  6$^+$  &   2$^+$ \\ \hline\noalign{\smallskip}
$E_{comp}^*$  & 1.35   & 1.76   & 1.43  &  1.33  \\
$E_{exp}$  & 1.52   & 2.75  &  3.19 &   1.03 \\
\noalign{\smallskip}\hline
\end{tabular}
\end{center}
\end{table}

We first calculate the $2^+$ states whose lowest energies, shown in
Table~\ref{tab4}, turn out to be $1.35$~MeV and $1.33$~MeV for $^{42}$Ca 
and $^{50}$Ca, respectively.  Obviously, different $2^+$ states can
also be achieved by couplings of a number of other than $f$ and $p$ partial 
waves, which implies that many $2^+$ excited valence states do actually exist,
as in fact experimentally observed \cite{fir99}. 
Proper comparison to measured spectra then requires detailed structure and
decay information, which is beyond the scope of the present paper.  We
only note here that the computed energies are similar to the energies of the
lowest excited 2$^+$ states present in the
measured spectra (1.52 MeV and 1.03 MeV, respectively \cite{fir99}).
For $^{42}$Ca, where only relative $f$-waves between the core and
the valence neutrons are included, we observe that the $2^+$ state
is almost fully dominated by the $(f_{7/2},f_{7/2})$ 
configuration, with an almost negligible contribution (less than 1\%)
from the $(f_{5/2},f_{5/2})$ and the $(f_{5/2},f_{7/2})$ components. For $^{50}$Ca, about 96\%
is provided by the $(p_{3/2},p_{3/2})$ configuration, whereas the remaining
4\% is provided by the $(p_{3/2},p_{1/2})$ component.

There is a number of possible $J^{\pi}$ states which can also be achieved by
valence excitation.  To show the capability and flavor of the unified
method, we shall here select a few of such states with the simplest
configurations.  For $^{42}$Ca we extend the sequence of $0^+$ and
$2^+$ to $4^+$ and $6^+$ states,  which are all achieved by the
core-neutron $f$-waves coupling.  The energies are also given in Table~\ref{tab4}.
They are only slightly bigger than the $2^+$ energy, and below the
experimental values of the lowest 4$^+$ and 6$^+$ states \cite{fir99}.
This similarity is perhaps not surprising, since the core is virtually unchanged (less
than a few keV variation).  Consequently the neutron-core potential is pretty
much the same for all angular momenta. In any case, it is important to take into
account that the numbers given in Table~\ref{tab4} have been obtained 
assuming for the 2$^+$, 4$^+$ and 6$^+$ states the same three-body force as
the one providing the correct two-neutron separation energy for the ground 0$^+$ state.

\begin{figure}
\begin{center}
\includegraphics[width=4cm,angle=0]{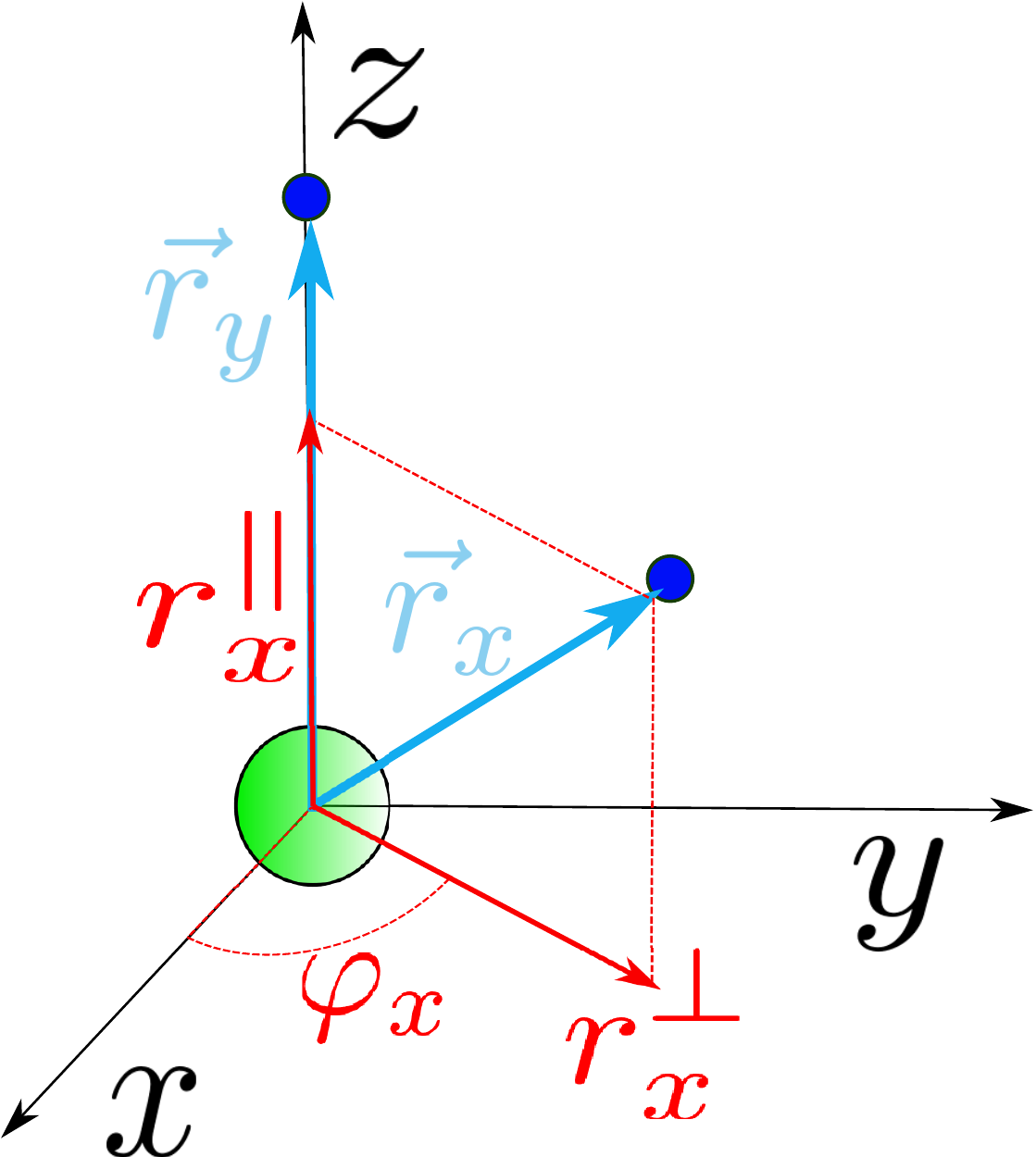}
\end{center}
\caption{ Definition of the intrinsic coordinates $r_x^\parallel$ and $r_x^\perp$ in a three-body system
where the $z$-axis is chosen along the Jacobi coordinate $\bm{y}$ defined between the
core and one of the valence neutrons. For $^{42}$Ca and $^{50}$Ca, whose respective cores 
are 40 and 48 times heavier than the neutron mass, the $\bm{y}$-Jacobi coordinate can be, to 
a large extent, considered as going from the core center of mass to the second valence neutron.}
\label{fig4} 
\end{figure}

From the wave functions it is also possible to inspect the structure of the different
$J^\pi$ states. We do it with the help of the intrinsic coordinates described in Fig.~\ref{fig4}, 
which are defined after choosing the $z$-axis along the $\bm{r}_y$-coordinate 
(proportional to $\bm{y}$-Jacobi coordinate) in the second or third Jacobi set.  In 
our case, where the core is 40 or 48 times heavier than the valence particle, the vector $\bm{r}_y$ can safely be assumed
to go from the center of mass of the core to one of the valence neutrons. We then define
$r_x^\parallel$ and $r_x^\perp$ as the projection of $\bm{r}_x$, which defines the position of the
second neutron, along $\bm{r}_y$ and
on the plane perpendicular to $\bm{r}_y$, respectively. Using now $r_x^\parallel$ and $r_x^\perp$
as coordinates, it is simple to visualize the position of the second valence neutron relative to the
first one. We do this with the help of the density function:
\begin{equation}
{\cal F}(r_x^\parallel,r_x^\perp)=\int r_x^\perp d\varphi_x d^3r_y 
\left|  \Psi_{3b} (\bm{x}, \bm{y}) \right|^2,
\label{eq4}
\end{equation}
where $ \Psi_{3b} (\bm{x}, \bm{y})$ is the solution of Eq.(\ref{eq2}). The function ${\cal F}$ is normalized to 1 
after integration over $r_x^\parallel$ and $r_x^\perp$.

\begin{figure}
\begin{center}
\includegraphics[width=\textwidth,angle=0]{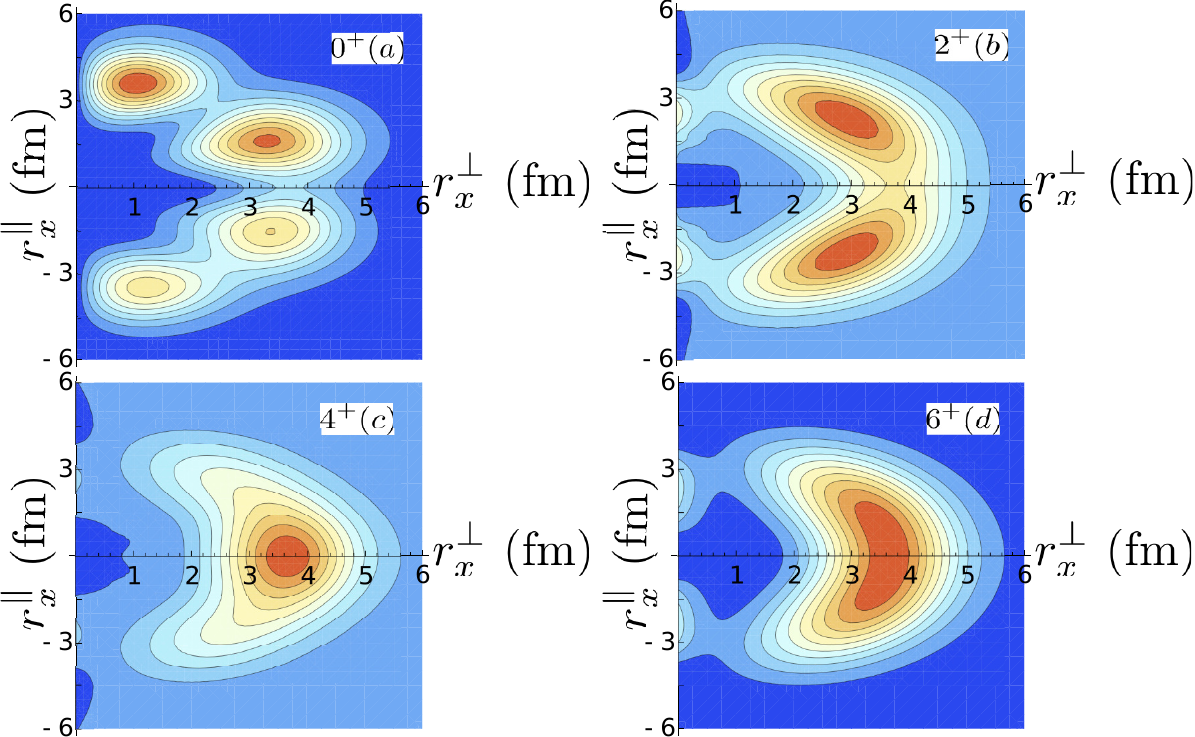}
\end{center}
\caption{Contour plots for the intrinsic structure, as defined by the function ${\cal F}(r_x^\parallel,r_x^\perp)$ in Eq.(\ref{eq4}),
for the $0^+$, $2^+$, $4^+$, and $6^+$ states in $^{42}$Ca. The figures have been obtained using the SKM$^*$ interaction. }
\label{fig5} 
\end{figure}

The contour plot of the density function (\ref{eq4}) is shown in Fig.~\ref{fig5} for the $0^+$, $2^+$, $4^+$, and $6^+$
states in $^{42}$Ca. The first thing we observe is the symmetry of the function along the $r_x^\parallel=0$ axis.
This is a result of the analytical structure of ${\cal F}(r_x^\parallel,r_x^\perp)$,
from which it is possible to see that when the three-body wave function $\Psi_{3b}$ does not mix even and 
odd core-neutron relative
orbital angular momenta, the equality ${\cal F}(r_x^\parallel,r_x^\perp)={\cal F}(-r_x^\parallel,r_x^\perp)$
is fulfilled. For this reason, since the $2^+$, $4^+$, and $6^+$ states have been obtained including
$f$-waves only, the corresponding density functions are perfectly symmetric along the $r_x^\parallel=0$ axis.
For the $0^+$ state, the almost negligible contribution from the $d$-waves is enough to slightly weaken 
the symmetry along the same axis. The result is that any maximum observed at an angle $\theta$ between 
$\bm{r}_x$ and $\bm{r}_y$ in Fig.~\ref{fig4} (an angle $\theta$ between the position vectors of the two valence neutrons) 
has a counterpart at an angle $\pi-\theta$.

As we can see in the figure, even if their excitation energies are not that different, the structure of the 
four computed states in $^{42}$Ca is very different. Four maxima are observed for the 0$^+$ state and 
two for the 2$^+$ state. For the $4^+$ and $6^+$ states we see that the second neutron has a preference
to stay in the plane perpendicular to the direction defined by the first neutron, but even in this case a difference
is seen, since in the 6$^+$ case a deviation from the perpendicular plane is clearly more likely.

\section{Summary and conclusions}

In this work we have extended applications of the unified method
described in Ref.\cite{hov18}.  The principal virtue of the method is
that short- and large-distance properties are directly connected, and
both treated in controlled and well-understood approximations. At the
moment short distances are described in the mean-field approximation,
whereas larger distances are described with state-of-the art methods.

The method assumes a clusterized structure for the nucleus, and the
internal many-body cluster degrees of freedom and the few-body
relative cluster structure are treated self-consistently.  The mean
field generated by a given cluster is distorted by the presence of the
others, while at the same time determining the cluster-cluster
interaction.  In other words, the force used in the mean-field calculations also
provides by folding the interaction between the clusters.  One of the
simplest clusterized systems is a many-body core surrounded by a
group of valence nucleons.  This structure is particularly suitable to
describe nuclei close to the proton and neutron driplines, as shown in
the previous works \cite{hov17,hov18b,hov18c}.

In the present work we apply the method to ordinary well-bound nuclei
far from the driplines.  The purpose is to investigate if the assumed
clusterized model structure forces the system to possess the
properties characteristic for dripline nuclei.  With this main purpose
in mind, we have chosen the $^{42}$Ca and $^{50}$Ca isotopes as an
illustration.  We have compared the results obtained after pure
Hartree-Fock mean field calculations and after use of the unified
method describing the system as a $^{40}$Ca or $^{48}$Ca core plus two
valence neutrons.

Four different nucleon-nucleon Skyrme interactions have been used for the
Hartree-Fock calculation of the core, as well as for the core-neutron
interaction obtained by folding.  The Hartree-Fock calculation is done
for both core and total nucleus, and the results are very similar for
the different interactions.  Also the results in the unified method
(core with mean-field and valence neutrons with a few-body technique)
only weakly depend on the choice of the Skyrme interaction.  The latter is
seen as the hyperspherical adiabatic potentials entering at the
three-body level are very similar, and therefore give rise to
correspondingly similar nuclear properties.

More specifically, we considered energies and radii as the
characteristic structure properties.  The comparison to pure
uncorrelated Hartree-Fock results are appropriate, since the unified
method introduces correlations in valence space.  The unified method
with the free-space interaction looses $10-15\%$ in the two-neutron
separation energy in comparison with pure Skyrme Hartree-Fock
calculations.  This is consistent with the typical results obtained
after a standard three-body calculation for dripline nuclei, where the
system usually is underbound when phase equivalent potentials are used
to exclude the Pauli forbidden states.  We attribute this result to
the combination of neutron-core Skyrme interactions and the unlimited
Hilbert space for the valence neutrons with the free-space
interaction.

In any case, this difference in the binding energy is not disturbing,
since the structure of the strongly bound as well as dripline nuclei
are improved.  The structure in the unified method remains basically
unchanged when the experimental two-neutron separation energy is
recovered with an effective three-body force.  The substantial benefit
is that the new treatment allows correlations between core and valence
particles.  The structure changes are seen by comparison of relative
sizes quantified as second radial moments obtained with the
Hartree-Fock and the unified methods.  The second moments of
two-nucleon distances are rather similar, but the valence distances
are significantly larger than both the Hartree-Fock sizes and the
corresponding pair distances between core nucleons.

After these ground state calculations we increased the angular
momentum to investigate excited states where the core is still angular
momentum $0^+$, but influenced by the resulting structure of the
valence neutrons. For $^{42}$Ca the $f_{7/2}$ neutron-core orbit is
by far the dominating valence configuration in the ground state, and $2^+$,
$4^+$ and $6^+$ states can be formed from the same orbits for two identical
nucleons. We approximate for simplicity to only $f$-partial waves.
For $^{50}$Ca the dominating valence configurations are $p$-waves,
where only $2^+$ states can be formed for identical particles.

The computed excitation energies are in reasonable agreement
with the experimental values for the lowest $2^+$ states, in both 
$^{42}$Ca and $^{50}$Ca, and the lowest $4^+$ and $6^+$ states 
for $^{42}$Ca , although an  unambiguous connection between the 
computed and measured values would require a more detailed analysis,
since many different partial wave configurations can provide the 
same total angular momentum.
We have also shown that the use of the unified method makes accessible 
the investigation of the spatial structure of the different states. In fact,
the computed states in $^{42}$Ca, although not very 
different in energy, show a very different spatial distribution. 

In conclusion, the clusterization assumption in the unified method
previously used to describe halo structure at driplines, is also
applicable to well-bound nuclei close to the valley of stability.
Although correlations are allowed in valence space these well-bound
structures are rather similar to those obtained when all core and
valence nucleons are treated equally.  To have more predictive power
for energies, it would be desirable with a better adjustment of the
valence interaction to the available space, see \cite{hov18} for a
discussion and an attempt of an answer.  This should include special
consideration of the pairing channel, which seems so obviously
connected to the interaction between the two valence neutrons.


\end{document}